\documentclass[sigconf, nonacm]{acmart}
\usepackage{tikz, pgfplots}
    \usepgfplotslibrary{colorbrewer}

\usepackage{siunitx}
\pgfplotsset{compat=1.18} 
\usepackage{tabularray}
\usepackage{csquotes}
\usepackage{graphicx}
\usepackage{xcolor}
\usepackage{xspace}
\usepackage{caption}
\usepackage{subcaption}

\newcommand{\eg}{e.\,g.\xspace}

\newcommand{\etal}{et~al.\xspace}
    
\copyrightyear{2023}
\acmYear{2023}
\setcopyright{acmlicensed}\acmConference[MEMSYS '23]{The International Symposium on Memory Systems}{October 2--5, 2023}{Alexandria, VA, USA}
\acmBooktitle{The International Symposium on Memory Systems (MEMSYS '23), October 2--5, 2023, Alexandria, VA, USA}
\acmPrice{15.00}
\acmDOI{10.1145/3631882.3631899}
\acmISBN{979-8-4007-1644-7/23/10}

\begin{document}

\title{QCEDA: Using Quantum Computers for EDA}

\author{Matthias Jung}
\email{m.jung@uni-wuerzburg.de}
\affiliation{%
  \institution{Fraunhofer IESE / JMU Würzburg}
  \city{Kaiserslautern / Würzburg}
  \country{Germany}
}

\author{Sven O. Krumke}
\email{sven.krumke@math.rptu.de}
\affiliation{%
  \institution{RPTU Kaiserslautern}
  \city{Kaiserslautern}
  \country{Germany}
}

\author{Christof Schroth}
\email{christof.schroth@iese.fraunhofer.de}
\affiliation{%
  \institution{Fraunhofer IESE}
  \city{Kaiserslautern}
  \country{Germany}
}

\author{Elisabeth Lobe}
\email{Elisabeth.Lobe@dlr.de}
\affiliation{%
  \institution{DLR}
  \city{Braunschweig}
  \country{Germany}
}

\author{Wolfgang Mauerer}
\email{wolfgang.mauerer@othr.de}
\affiliation{%
  \institution{OTH Regensburg}
  \city{Regensburg}
  \country{Germany}
}

\begin{abstract}
The field of \textit{Electronic Design Automation} (EDA) is crucial for microelectronics, but the increasing complexity of \textit{Integrated Circuits} (ICs) poses challenges for conventional EDA: Corresponding problems are often NP-hard and are therefore in general solved by heuristics, not guaranteeing optimal solutions. Quantum computers may offer better solutions due to their potential for optimization through entanglement, superposition, and interference. Most of the works in the area of EDA and quantum computers focus on how to use EDA for building quantum circuits. However, almost no research focuses on exploiting quantum computers for solving EDA problems. Therefore, this paper investigates the feasibility and potential of quantum computing for a typical EDA optimization problem broken down to the Min-$k$-Union problem. The problem is mathematically transformed into a \textit{Quadratic Unconstrained Binary Optimization} (QUBO) problem, which was successfully solved on an IBM quantum computer and a D-Wave quantum annealer.
\end{abstract}

\maketitle

\section{Introduction}
As one of the most important areas of microelectronics, \textit{Electronic Design Automation} (EDA) has a long history, dating back to the mid-1960s. Nevertheless, EDA methods are still being intensively developed with the inclusion of the latest algorithms and technologies. In recent years, with the development of semiconductor technology, the complexity of \textit{Integrated Circuits} (IC) has increased exponentially, posing challenges to the scalability and reliability of circuit design. Therefore, EDA algorithms and software need to be more effective and efficient to handle an extremely large search space with low runtime. However, a large number of the problems in EDA, such as placement and wiring or scheduling, are NP-hard. Therefore, there is no algorithm for conventional computers that can solve these problems efficiently, i.e., in polynomial time depending on the problem size. Rather, the processing time grows exponentially. This means in the worst case that, for large but still reasonable problem sizes, a classical computer might have to compute millions of years to find an optimal solution, and that this situation cannot be relaxed by simply improving the performance of classical computers.
Thus, in practice, these problems can only be solved with the help of approximation algorithms or heuristics, which find feasible solutions but in general do not provide the mathematically optimal result. Quantum computers can take advantage of entanglement, superposition, and interference to speed up optimization algorithms through massive parallelism. Thus, for the EDA problems, there is the potential to achieve a significant speedup compared to a classical computer.  

Most of the works in the area of EDA and quantum computers focus on how to use EDA for building quantum circuits. However, to the best of our knowledge, almost no research focuses on exploiting quantum computers for EDA problems.
A typical EDA optimization problem is presented by the authors of~\cite{junhei_16, natjun_20}. The objective is to discover an optimal address mapping of a specific application for a \textit{Dynamic Random Access Memory} (DRAM), which is composed of banks, rows, and columns. This mapping is typically achieved through a hardware scrambler in the memory controller. The aim of the EDA problem is to determine an optimal configuration for this hardware scrambler, reducing the number of row misses, thereby increasing bandwidth and reducing latency. It has been shown by~\cite{junhei_16,natjun_20} that this problem is NP-hard and that the core of the problem can be reduced to the so-called Min-$k$-Union problem. 

In this paper we investigate the feasibility and discuss potential of quantum computing for this specific EDA optimization problem. In order to speedup the calculation, the goal of this paper is to formulate the Min-$k$-Union problem for the quantum computer. 
While we find that currently available quantum computer prototypes do not scale to realistically sized problem instances, we quantitatively estimate required machine sizes,
and verify general feasibility of our approach on an IBM quantum computer and a D-Wave quantum annealer,
and discuss paths towards quantum advantage on EDA.

In summary this paper makes the following contributions:

\begin{itemize}
    \item We show, for the best of our knowledge, for the first time, how an very specific EDA problem can be formulated to be executed on a quantum computer.
    \item In order to achieve that, we present, for the first time, a \textit{Quadratic Unconstrained Binary Optimization} (QUBO) formulation of the Min-$k$-Union problem.
    \item We execute this EDA problem on real quantum computers and prove the feasibility of this approach.
    \item We show how this problems scales for real world problem instances and point out limitations for the future.
\end{itemize}

The paper is structured as follows: Section~\ref{sec:rel} discusses the related work.
The mathematical description and the transformation for the quantum computers of the Min-$k$-Union problem is discussed in Section~\ref{sec:math}. The results on the execution on two real quantum machines is presented in Section~\ref{sec:results}. Finally, the paper is concluded in Section~\ref{sec:conclusion}.

\section{Related Work}
\label{sec:rel}
Most of the works in the area of EDA and quantum computers focus on how to use EDA for building quantum circuits~\cite{soemeu_20}. 
For instance, the authors of~\cite{zulwil_18} present a logic synthesis for reversible circuits. Hillmich~\etal present new approaches for quantum circuit simulations based on decision diagrams~\cite{hilzul_22}. The synthesis and mapping of quantum circuits to specific hardware is presented in~\cite{zulpal_18, niedat_16, shebul_05, soemeu_20}. There exists also some work in the field of quantum circuit verification~\cite{burwil_21, burray_20}. 
However, to the best of our knowledge, so far, there exists no work with focus on exploiting quantum computers for specific EDA problems, although the potential for the other direction of this symbolic relationship of quantum computing and EDA has been highlighted by Raghunathan and Stok~\cite{ragsto_20}.
In~\cite{sutu_16} the authors analyze a quantum annealing approach to solve SAT problems. Like a lot of combinatorial optimization problems, the Min-$k$-Union problem could also be transferred to SAT and then further processed with the existing approaches. This transformation however introduces overhead in terms of variables and quadratic terms, which is why direct approaches are preferable.

Quantum computing in general and quantum optimization
in particular have seen a large body of work come into existence
during the last years, yet many aspects are not yet 
fully understood. In particular, any quantum processing
units (QPUs) that are available either commercially or in
research labs today suffer from considerable imperfections
and resource constraints, and are therefore termed 
\emph{noisy, intermediate-scale quantum} (NISQ) machines.
This influences both, the choice of the optimization algorithm
and the approach to evaluation.

\emph{Variational quantum algorithms}, that are particularly 
tailored to the capabilities NISQ-era machines, include the
QAOA family of algorithms (see, e.g.,~Refs.~\cite{Alam:2020,Wang:2021}). 
These algorithms aim at solving optimization problems and are hypothesized to achieve computational gains over classical approaches, albeit a 
practical advantage has not yet been observed to the best of our knowledge in any field. Nonetheless, it has been shown that
it is impossible for any generic classical algorithm to 
efficiently sample the output distribution of QAOA algorithms,
even in very restricted scenarios (i.e., with the level parameter $p=1$, which we elaborate in Section~\ref{sec:opti}),
at least when generally accepted 
complexity-theoretic assumptions are true~\cite{Farhi:2016}. While
this indicates quantum advantage in a certain sense, further experimental progress
is required to explore the capabilities of the approach in
relevant scenarios, in particular when executed on noisy devices. 
Leymann~\etal~\cite{Leymann:2020} discuss the (considerable)
impact of imperfections in NISQ machines on quantum algorithms;
Greiwe~\etal~\cite{Greiwe:2023} show illustratively the performance degradation
of typical quantum algorithms under the influence of noise.
How to benchmark quantum algorithms is considered, amongst others,
by Becker~\etal~\cite{Becker:2022}, Tomesh~\etal~\cite{Tomesh:2022} and
Resch~\etal~\cite{Resch:2021}.
\section{Formulation for Quantum-Based Optimization}\label{sec:math}
In this section, we present the details of our quantum formulation of the EDA problem. Given that QC is a relatively new paradigm, it behooves to first recall some fundamentals on how QPUs operate algorithmically, as this differs substantially from the patterns known from classical computing. We also discuss the primitives available for our formulation and provide a rationale for our choice of empirical evaluation approach.
\subsection{Quantum Optimization}\label{sec:opti}
Multiple quantum approaches allow us to solve our subject
problem; two are particularly common for currently available machines:

(1) The \emph{Quantum Approximate Optimization Algorithm}
(QAOA)~\cite{Farhi:2014} is an iterative, hybrid
quantum-classical algorithm for \emph{gate-based} QPUs that can be
used to seek minimal solutions to QUBO problems. Roughly speaking, QAOA applies a set of parameterized quantum operations
including an evaluation of the target function
to an initial state, samples the resulting probability distribution
of possible outcomes caused by quantum superposition, and
then uses classical optimization to update parameters for the quantum
operations that lead to improved measurement results in the next iteration.
Additionally, the core quantum part of the algorithm can be
performed \(p\) times in each iteration, correspondingly increasing
the number of parameters. For perfect QPUs, it can be shown that results 
improve with increasing \(p\) and thus increased computational effort,
whereas NISQ machines will experience a trade-off between a more expressive computation
and increasing amounts of noise with increasing~\(p\).

Possibilities to improve the performance of QAOA on NISQ
machines are plentiful: Noise mitigation techniques 
(\eg,~\cite{Berg:2022,Lao:2022}); choosing good initial parameters (often referred to as warm-starting) by classical 
(\eg,~\cite{Egger:2021}) and machine learning approaches 
(\eg,~\cite{Khairy:2020}); by reducing classical
optimization to lower-dimensional, nearly equivalent spaces
(see, \eg, Ref.~\cite{Zhou:2020}).
Note that recent insights on variational quantum circuits
in general and QAOA in particular (for instance, using
large Fourier series~\cite{Landman:2022}) give criteria for
the feasibility of classically approximating quantum variational 
algorithms, which limit the potential of quantum approaches.
Likewise, the detrimental impact of noise on QAOA
has been characterised experimentally (\eg,~\cite{Harrigan:2021}, based
on a sound theoretical understanding (\eg,~\cite{Marshall:2020,Xue:2021}), which
further limits the merit of evaluations on current-generation hardware.

(2) \emph{Quantum annealing}\footnote{Quantum annealing is
a restricted variant of the more general adiabatic evolution of a quantum system. In turn, QAOA can be seen as a finite approximation to an adiabatic evolution, which is recovered in the limit \(p\rightarrow\infty\). Consequently, most of the remarks on the need for empirical evaluation of the performance of NISQ machines on our subject problem apply in equal measure to both approaches.} (respectively adiabatic quantum computation~\cite{Albash:2018}) is -- depending on the point of view -- a particular transformation executed on a quantum computer
using global operations~\cite{Wintersperger:2023}, or is performed by a special class of
machines purpose-built~\cite{McGeoch:2020} to solve, respectively, approximate~\cite{Sax:2020} QUBO problems. The scheme operates similar to classical annealing procedures, yet it can benefit from
quantum effects to speed up the underlying optimization problem~\cite{McGeoch:2014}.
For the physical and algorithmic details of these base patterns, and other algorithmic possibilities, we refer to the available introductory texts on quantum computing~\cite{Nielsen:2000}, or recent reviews~\cite{Bharti:2022}.

We emphasize that the focus of our paper is to introduce the required reformulation of the EDA problem, which can be used as starting point for all aforementioned approaches. 
Since there is no unified theoretical understanding behind
all variants of quantum optimization discussed above, it is challenging to predict which variant is best suited to a given
combination of machine and problem, and an comprehensive empirical evaluation
is mandated. However, for realistic settings, this necessitates
a comparison with classical heuristics and probabilistic approaches, especially regarding to their average-case performance.
The complexity of this task is universally appreciated,
independent of quantum computing, and considered at textbook level 
(see, \eg,~\cite{Arora:2006}). Furthermore, quantum performance 
evaluation itself is highly non-trivial~\cite{McGeoch:2019}.
Since NISQ machines fail to provide the required 
qubit resources for realistic instances of our problem
by a wide margin (cf.~Tab.~\ref{tab:qbits}), we deliberately
refrain from conducting an empirical performance evaluation
beyond the scale of toy problems in this paper.
Finally, note that it would be possible to derive runtime 
bounds for a given task for the mechanism underlying QAOA 
and annealing, which unfortunately requires knowledge of the
so-called \emph{minimal spectral gap}, which is
as hard to compute as solving the problem itself. 

\subsection{Problem Extraction}
DRAMs consist of memory cells organized into memory arrays composed of columns, rows, and banks. The amalgamation of primary and secondary sense amplifiers within a bank's memory arrays is commonly termed a \textit{row buffer}. Typically, the row buffer possesses a capacity ranging from $1\mbox{KB}$ to $8\mbox{KB}$, which is known as the DRAM page size. It operates as a compact cache, storing the most recently accessed row within the bank. The latency of a memory access to a bank is heavily influenced by the state of this row buffer. A memory access targeting the same row as the one currently cached in the buffer (referred to as a \textit{row hit}) results in minimal latency and energy consumption. Conversely, if a memory access targets a different row than the one stored in the buffer (referred to as a \textit{row miss}), it leads to heightened latency and energy consumption. Meanwhile, the concurrent access of activated rows in distinct banks without penalty, known as \textit{Bank Parallelism}, can be harnessed to enhance overall performance.
Thus, the achieved DRAM bandwidth and latency strongly depends on the access patterns of the applications. Therefore, memory controllers have configurable address scramblers, which permute the address bits by means of simple lookup tables or a network of multiplexers, in order to maximize the sustainable DRAM bandwidth. 

The EDA problem in focus is to find an optimal configuration for the scrambler such that row misses are minimized and the bank parallelism is maximised. The work presented in~\cite{natjun_20} demonstrated that a multi-bank DRAM can be effectively simplified into a single-bank DRAM, given the independent operation of all DRAM banks. 
Consequently, we will exclusively focus on DRAMs with a single bank for the remainder of this paper. It has been shown by~\cite{junhei_16,natjun_20} that this problem is NP-hard and that the core of the problem can be reduced to the so-called Min-$k$-Union problem. 
As the solution to this problem holds the highest time-criticality, our primary focus lies in accelerating its resolution.

Roughly summarizing the deductions of~\cite{natjun_20}, the problem can be extracted as follows: 
We are given a memory address sequence for an arbitrary application. 
These addresses should be mapped to a DRAM memory, 
where the goal is to find an assignment of the address bits to new row and column bits -- and in the general case also to bank bits -- 
such that the number of row misses is minimized.
A row miss appears wherever we have at least one bit flip from one address to the following in the assigned row bits. Due to the resulting overhead, this should happen as few as possible.
Therefore, each column of the stacked addresses defines a set
containing the row numbers where a bit change appears 
and the goal is to select a specified number of these sets 
where we have the least row misses, i.e., the least number of elements. 

The resulting problem is the Min-$k$-Union problem,
which is defined mathematically as this: Given a finite ground set~$V$, a collection $\mathcal{S}\subseteq 2^V$ of subsets of the ground set (where $2^V$ denotes the power set of~$V$) and  $k\in\mathbb{N}$, the goal is to choose exactly $k$~sets $M_1,\dotsc,M_k\in\mathcal{S}$ such that the cardinality of the union $T=\bigcup_{i=1}^k M_i$ of the chosen sets is as small as possible.
As shown in Theorem 1 of~\cite{natjun_20}, finding an optimal permutation, i.e., an optimal assignment from address bits to row bits, is equivalent in solving an instance of the Min-$k$-Union problem where $k$ denotes the number of row bits that have to be assigned.
The deduction to this problem will also become more clear with the concrete example explained in Section~\ref{sec:example}.

\subsection{QUBO Formulation}
The Min-$k$-Union problem is known to be NP-hard to solve and even NP-hard to approximate~\cite{chldin_17}.  Thus, it seems unlikely that one is able to find an algorithm which solves all instances efficiently in polynomial time (since this would imply that the complexity classes P and NP coincide). Moreover, essentially all known formulations of the Min-$k$-Union problem as \emph{linear} integer programs suffer from the weakness that the corresponding linear relaxation is rather weak, meaning that integer linear programming solvers tend to explore many nodes in the branch-and-bound tree which in turn means a rather inefficient solution procedure.

Thus, the approach taken in this paper is different: Instead of using a \emph{linear} formulation for the Min-$k$-Union problem, we use a \emph{quadratic} formulation, which is suitable for quantum computing. As mentioned before, one type of optimization problems which have proven to be appropriate in this respect are \emph{Quadratic Unconstrained Binary Optimization} problems (QUBOs). A QUBO is an optimization problem of the form
\begin{align*}
    \min \quad & \sum_{i=1}^n \sum_{j=1}^n q_{ij} x_i x_j = x^T Q x=H(x)\\
    \text{s.t.} \quad &x\in\{0,1\}^n,
\end{align*}
where $Q=(q_{ij})_{i,j=1,\dotsc,n}$ is a given matrix. In the following, we will show how to transform the Min-$k$-Union problem to a QUBO.


In order to formulate the Min-$k$-Union problem as a QUBO, we define binary variables with the following meaning: For $M\in \mathcal{S}$ we set the binary variable
\begin{align*}
    x_M=\begin{cases}
        1, & \text{if $M\in\mathcal{S}$ is chosen},\\
        0, & \text{otherwise}.
    \end{cases}
\end{align*}
We also have binary variables~$y_v\in \{0,1\}$  with the following meaning:
\begin{align*}
    y_{v}=\begin{cases}
        1, & \text{if $v$ is contained the union }\\
        & \text{of the chosen $k$~sets},\\
        0, & \text{otherwise}.
    \end{cases}
\end{align*}
We now construct the objective~$H$ of the QUBO, which is
is composed of three parts, i.e., $H=H_A+H_B+H_C$, each of which is non-negative, and which we describe now.

We first have
\begin{align}\label{eq:1}
  H_A(x) :=A\left(k-\sum_{M\in\mathcal{S}} x_M\right)^2\geq 0,
\end{align}
where $A>0$ is a constant to be chosen later. Obviously $H_A(x)=0$ if and only if the selection of sets described by $x$ contains exactly $k$~sets. The term $H_A$ is intended as a ``penalty term'' and we will show at the end of this section how to determine the penalty parameter~$A>0$, such that any optimal solution $x^*$ of the QUBO fulfils $H_A(x^*)=0$, i.e., forms a feasible solution of the original problem.

In order to properly count the number of elements
in the union $\bigcup_{M\in \mathcal{S}: x_M=1} M$ of the chosen sets, we need to ensure the activation of an element $v$ once a set is activated in which contains the element. This can be done with the inequality constraint $y_v\geq x_M$ for each $v\in V$ and each $M\in\mathcal{S}$ with $v\in S$. For the QUBO reformulation, consider the term $t_{M,v}=(1-y_v)x_M\geq 0$ for $M\in\mathcal{S}$ and $v\in M$.  If $x_M=1$, then the only way to achieve $t_{M,v}=0$ is to set $y_v=1$. If in turn $x_M=0$, then $t_{v,M}=0$ no matter what the value of~$y_v$ is. These considerations lead us to our new part~$H_B$, where $B>0$ is again a suitable penalty parameter to be determined later:
\begin{align}\label{eq:4}
 H_B(x, y) &:= B \sum_{v\in V}\sum_{M\in\mathcal{S}:v\in M} \left(1-y_v\right)x_M.
\end{align}

The third term~$H_C$ is the actual objective function of the Min-$k$-Union problem:
\begin{align}\label{eq:5}
    H_C(y) :=C\sum_{v\in V} y_{v}.
\end{align}
We now address the choice of the constants $A$, $B$ and $C$ in the above formulation.  Recall that $H_A\geq A$ if our choice of sets does not contain exactly $k$~sets. If we have set $x_M=1$ and $y_v=0$ for some element $v\in M$, then $H_B\geq B$.  Furthermore, we always have $0\leq H_C\leq C|V|$.  Thus, if we choose $A=B>C|V|$, then any solution that minimizes $H=H_A+H_B+H_C$ will have $H_A=H_B=0$ and thus form a feasible solution to the Min-$k$-Union problem, where $H_C$ correctly counts the number of chosen elements.  In particular $C=1$, $A=B=|V|+1$ satisfy this condition.

This means, in the end, the formulation given above integrates all constraints of the Min-$k$-Union problem via~\eqref{eq:1} and~\eqref{eq:4} into the objective in an \emph{exact} formulation: Any optimal solution of the QUBO is in fact an optimal solution for the given instance of the Min-$k$-Union problem and one does not need to vary penalty parameters. The QUBO then is an \emph{unconstrained} problem which can be given to a quantum computer without any further manipulations.

Furthermore, this QUBO formulation has a number of advantages.  First, it uses only a number of variables, which is linear in the number of elements and sets. Due to their structure, the constraints of the original Min-$k$-Union problem formulation do not introduce any additional variables in the QUBO formulation.
This is important since the number of variables translates directly into the number of quantum bits needed.
Additionally, the coefficients of the resulting QUBO, i.e., the values $q_{i,j}$, have a simple structure; they are integer and only dependent on the parameter $k$ and the constants $A$, $B$ and $C$. This might support finding the solution using NISQ devices.

%
%
%
\begin{table}
	\caption{Sequence of Memory Addresses with Highlighted Bit-Toggling}
	\centering
	\begin{tabular}{|c|ccccc|}
		\hline
		$\mathbf{a}$& $a_{i,0}$ & $a_{i,1}$ & $a_{i,2}$ &  $a_{i,3}$ &  $a_{i,4}$ \\
		\hline
		$a_0$ & $1$ & $0$ & $0$ &  $0$ & $1$\\
		$a_1$ & $\pmb{0}$ & $0$ & $0$ &  $\pmb{1}$ & $1$\\
		$a_2$ & $0$ & $0$ & $\pmb{1}$ &  $1$ & $\pmb{0}$ \\
		$a_3$ & $0$ & $\pmb{1}$ & $1$ &  $\pmb{0}$ & $\pmb{1}$\\
		$a_4$ & $0$ & $1$ & $1$ &  $\pmb{1}$ & $1$ \\
		$a_5$ & $0$ & $1$ & $1$ &  $\pmb{0}$ & $1$ \\
		$a_6$ & $\pmb{1}$ & $1$ & $\pmb{0}$ &  $0$ & $\pmb{0}$  \\ 
		$a_7$ & $1$ & $1$ & $0$  & $0$ & $\pmb{1}$   \\
		$a_8$ & $1$ & $\pmb{0}$ & $\pmb{1}$  & $0$ & $1$   \\
		\hline
	\end{tabular}
	\label{tab:seq_example}
\end{table}
\section{Case Study}
To demonstrate the feasibility of our approach, we transform the example of~\cite{natjun_20} into a QUBO problem, employing the formalism presented in Section~\ref{sec:math}. First, we describe the artificial example. Second, we describe the execution on the IBM quantum computer and the D-Wave quantum annealer, and third, we discuss the results with respect to scalability of the approach. 
\label{sec:results}

\subsection{Example Problem}\label{sec:example}
Table~\ref{tab:seq_example} shows the memory address sequence $a$ for an artificial application. The addresses of this application shall be mapped in an artificial DRAM with 
$8$ rows ($3$ row address bits) and $4$ columns per row ($2$ column address bits). The goal is to find a selection of the address bits to serve as the row bits, such that the number of row misses is minimized. As mentioned before, the reduction to a single bank is suitable and we therefore assign no bank address bits here. 

The bold numbers in Table~\ref{tab:seq_example} represent the address bits that toggle between consecutive accesses. This toggling behavior is observed column-wise. Consequently, for each column $a_{*,j}$, $j \in {0,\ldots,4}$, we define a set $M_j$, which contains row indices $i \in {0,\ldots,8}$ corresponding to bit changes from $a_{i,j-1}$ to $a_{i,j}$, i.e., the highlighted numbers in the table. In this specific case, we have $M_0 = \{1,6\}$, $M_1= \{3,8\}$, $M_2 = \{2,6,8\}$, $M_3 = \{1,3,4,5\}$, and $M_4=\{2,3,6,7\}$. The size of set $M_j$ reflects the number of bit changes for the $j$-th address bit.

\begin{figure}
     \centering
     \begin{subfigure}[b]{1.0\linewidth}
         \centering
         \includegraphics[width=0.45\linewidth]{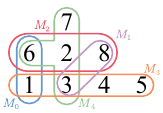}
         \caption{Example Problem}
         \label{fig:hyper:1}
     \end{subfigure}
     \hfill
     \begin{subfigure}[b]{0.45\linewidth}
         \centering
         \includegraphics[width=1.0\linewidth]{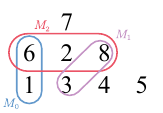}
         \caption{Solution-A}
         \label{fig:hyper:2}
     \end{subfigure}
     \begin{subfigure}[b]{0.45\linewidth}
         \centering
         \includegraphics[width=1.0\linewidth]{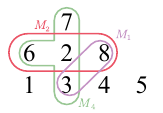}
         \caption{Solution-B}
         \label{fig:hyper:3}
     \end{subfigure}
        \vspace{10pt}
        \caption{Hypergraphs and Solutions for Example Problem Corresponding to Table~\ref{tab:seq_example}}
        \label{fig:hyper}
\end{figure}

Figure~\ref{fig:hyper:1} shows the sets which are formed. Since our artificial DRAM has 8 rows, we want to choose $3$ of these sets. 
A valid optimal solution of this small artificial example is easy to find: consider the union of the sets $M_0, M_1$ and $M_2$ given by $T := M_0 \cup M_1 \cup M_2 = \{1,2,3,6,8\}$, as shown in Figure~\ref{fig:hyper:2}. The elements of $T$ are exactly the positions of bit changes happening combined in columns $0, 1$ and $2$ and the size $|T|=5$ equals the minimal number of row misses.

\subsection{Execution on IBM Quantum Computer}
In this study, we have utilized Qiskit, an open-source python based SDK developed by IBM. Qiskit allows users to work with quantum computers at the circuit, pulse, and algorithm level. We formulated the example above as QUBO with Qiskit and executed it on the \texttt{ibmq\_ehningen} quantum computer, which is one of the IBM Quantum Canary Processors. Qiskit compiles the formulation in python in an according quantum circuit consisting of quantum gates. To ascertain reproducibility of our results~\cite{Mauerer:2022}, we will upload our code on Github once the publication is accepted.
The \texttt{ibmq\_ehningen} quantum computer has a total number of 27 qubits, where for our example only 13 qubits are required (8 nodes and 5 sets). It has a coherence time around 150 us.

The solution that the \texttt{ibmq\_ehningen} quantum computer found is shown in Figure~\ref{fig:hyper:3}. In contrary to our example solution in Figure~\ref{fig:hyper:2}, the quantum computer decided to pick the sets $M_1$, $M_2$ and $M_4$. However, this is a valid optimal solution as well, because the number of row misses (the number of Elements in the set $T= M_1 \cup M_2 \cup M_4$) is also 5. This proves that the execution of the EDA problem is feasible on a real quantum computer. 
\subsection{Execution on D-Wave Quantum Annealer}
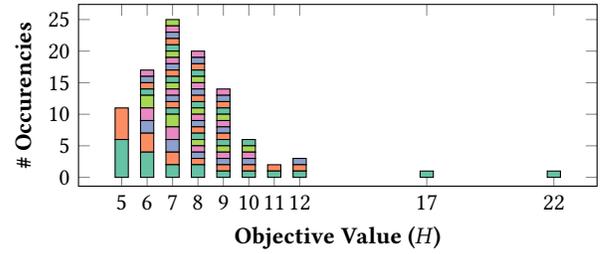
\begin{figure}
    \centering
     \begin{tikzpicture}
    \begin{axis}[
        ybar stacked,
        bar width=5pt,
        ylabel={\textbf{\# Occurencies}},
        xtick=data,
        ytick={0,5,10,15,20,25},
        xlabel={\textbf{Objective Value ($H$)}},
        height=4cm,
        width=\linewidth,
        cycle list/Set2-5,
        ]
        \addplot+[ybar, fill, draw=black] plot coordinates {
            (-13.0+18, 6)
            (-12.0+18, 4)
            (-11.0+18, 2)
            (-10.0+18, 2)
            ( -9.0+18, 1)
            ( -8.0+18, 1)
            ( -7.0+18, 1)
            ( -6.0+18, 1)
            ( -1.0+18, 1)
            (  4.0+18, 1)
        };
          
        \addplot+[ybar, fill, draw=black] plot coordinates {                
            (-13.0+18, 5)
            (-12.0+18, 3)
            (-11.0+18, 2)
            (-10.0+18, 1)
            ( -9.0+18, 1)
            ( -8.0+18, 1)
            ( -7.0+18, 1)
            ( -6.0+18, 1)
        };
        
        \addplot+[ybar, fill, draw=black] plot coordinates {
            (-13.0+18, 0)
            (-12.0+18, 2)
            (-11.0+18, 2)
            (-10.0+18, 1)
            ( -9.0+18, 1)
            ( -8.0+18, 1)
            ( -6.0+18, 1)
        };
         
        \addplot+[ybar, fill, draw=black] plot coordinates {
            (-13.0+18, 0)
            (-12.0+18, 2)
            (-11.0+18, 2)
            (-10.0+18, 1)
            ( -9.0+18, 1)
            ( -8.0+18, 1)
        };
        
        \addplot+[ybar, fill, draw=black] plot coordinates {
            (-13.0+18, 0)
            (-12.0+18, 2)
            (-11.0+18, 2)
            (-10.0+18, 1)
            ( -9.0+18, 1)
            ( -8.0+18, 1)
        };
        
        \addplot+[ybar, fill, draw=black] plot coordinates {
            (-13.0+18, 0)
            (-12.0+18, 1)
            (-11.0+18, 1)
            (-10.0+18, 1)
            ( -9.0+18, 1)
            ( -8.0+18, 1)
        };
                 
        \addplot+[ybar, fill, draw=black] plot coordinates {
            (-13.0+18, 0)
            (-12.0+18, 1)
            (-11.0+18, 1)
            (-10.0+18, 1)
            ( -9.0+18, 1)
        };
                 
        \addplot+[ybar, fill, draw=black] plot coordinates {
            (-13.0+18, 0)
            (-12.0+18, 1)
            (-11.0+18, 1)
            (-10.0+18, 1)
            ( -9.0+18, 1)
        };
                 
        \addplot+[ybar, fill, draw=black] plot coordinates {
            (-13.0+18, 0)
            (-12.0+18, 1)
            (-11.0+18, 1)
            (-10.0+18, 1)
            ( -9.0+18, 1)
        };

        \addplot+[ybar, fill, draw=black] plot coordinates {
            (-13.0+18, 0)
            (-12.0+18, 0)
            (-11.0+18, 1)
            (-10.0+18, 1)
            ( -9.0+18, 1)
        };
                 
        \addplot+[ybar, fill, draw=black] plot coordinates {
            (-13.0+18, 0)
            (-12.0+18, 0)
            (-11.0+18, 1)
            (-10.0+18, 1)
            ( -9.0+18, 1)
        };
                 
        \addplot+[ybar, fill, draw=black] plot coordinates {
            (-13.0+18, 0)
            (-12.0+18, 0)
            (-11.0+18, 1)
            (-10.0+18, 1)
            ( -9.0+18, 1)
        };
                 
        \addplot+[ybar, fill, draw=black] plot coordinates {
            (-13.0+18, 0)
            (-12.0+18, 0)
            (-11.0+18, 1)
            (-10.0+18, 1)
            ( -9.0+18, 1)
        };
                 
        \addplot+[ybar, fill, draw=black] plot coordinates {
            (-13.0+18, 0)
            (-12.0+18, 0)
            (-11.0+18, 1)
            (-10.0+18, 1)
            ( -9.0+18, 1)
        };
                 
        \addplot+[ybar, fill, draw=black] plot coordinates {
            (-13.0+18, 0)
            (-12.0+18, 0)
            (-11.0+18, 1)
            (-10.0+18, 1)
        };
                 
        \addplot+[ybar, fill, draw=black] plot coordinates {
            (-13.0+18, 0)
            (-12.0+18, 0)
            (-11.0+18, 1)
            (-10.0+18, 1)
        };
                 
        \addplot+[ybar, fill, draw=black] plot coordinates {
            (-13.0+18, 0)
            (-12.0+18, 0)
            (-11.0+18, 1)
            (-10.0+18, 1)
        };
                 
        \addplot+[ybar, fill, draw=black] plot coordinates {
            (-13.0+18, 0)
            (-12.0+18, 0)
            (-11.0+18, 1)
            (-10.0+18, 1)
        };
                 
        \addplot+[ybar, fill, draw=black] plot coordinates {
            (-13.0+18, 0)
            (-12.0+18, 0)
            (-11.0+18, 1)
            (-10.0+18, 1)
        };
                 
        \addplot+[ybar, fill, draw=black] plot coordinates {
            (-13.0+18, 0)
            (-12.0+18, 0)
            (-11.0+18, 1)
        };
        
    \end{axis}
    
 \end{tikzpicture}
    \caption{Histogram of solutions from D-Wave run}
    \label{fig:histogram}
\end{figure}

Over D-Wave's cloud platform \textit{Leap}\footnote{\url{https://cloud.dwavesys.com/leap/}}, using the trial access, and their python library \texttt{dwave-ocean-sdk}\footnote{\url{https://github.com/dwavesystems/dwave-ocean-sdk}}, one can easily send problems to the connected quantum annealers of the current \textit{Advantage} generation with about 5000 qubits. 
Using the internally implemented embedding and de-embedding strategy, we were able to submit the above test instance in our QUBO formulation directly. The size of the annealing sample was set to 100 and besides that we used the default solver parameters, such as 20 $\mu s$ annealing time. The full sample set for a single run is shown in Figure~\ref{fig:histogram}. 
The different parts of the bars indicate different solutions yielding the same objective value. We obtained both optimal solutions at once, where Solution~A was found in 6 of the 100 cases and Solution~B in 5 cases. The other cases are sub-optimal solutions with varying objective values. Note that the values differ slightly in subsequent runs, due to the heuristic nature of the machines. 
The implementation of the corresponding test script was supported by the tool \texttt{quark}\footnote{\url{https://gitlab.com/quantum-computing-software/}}. 
\subsection{Discussion}
\begin{table}[]
    \centering
    \caption{Required Number of Qubits, Benchmarks from~\cite{nat_19}}
    \label{tab:qbits}
    \begin{tabular}{llll}
         Benchmark & Elements & Sets & Qubits  \\\hline
         filter7   & 524288   & 19   & 524307 \\
         rot6      & 65536    & 16   & 65552  \\
         rot3d7    & 2097152  & 21   & 2097173\\
         NN8       & 356400   & 22   & 356422 \\\hline
    \end{tabular}
\end{table}
Having demonstrated the feasibility of executing the formulation on two quantum machines, the question that arises is how well the formulation scales for real-world applications that store their data in real DRAM-Chips like DDR5 oder LPDDR5. As previously mentioned, the number of required qubits depends on the sum of elements and sets, thus scaling linearly. Table~\ref{tab:qbits} illustrates the necessary number of qubits for various real-world benchmark applications~\cite{nat_19}. The primary factor influencing the qubit requirement is the number of elements. Since the number of elements corresponds to the number of unique DRAM addresses present in a benchmark, it serves as a reasonable approximation for the required number of qubits.
Let us highlight that the problem size only grows linearly with the number of elements and sets, compared to other approaches that exhibit non-linear growth in problem size.
Given this favorable property, we believe that solving the EDA 
problem using quantum computers or quantum accelerators in the future holds promise.

Of course, the scope of our empirical feasibility evaluation 
remains far from touching practical utility.
While today's quantum computers lack a sufficient number of qubits to solve real-world instances, we are currently witnessing exponential growth in qubit availability, backed by ambitious
roadmaps of commercial vendors. 
Additionally, quantum-inspired computational accelerators like 
Fujitsu's digital annealer will allow us, pursuing a slightly
different route, to explore considerably 
larger instances in future work.
We therefore believe our 
approach to contribute an important milestone towards using quantum  computers for EDA workloads, as it 
establishes a solid foundation for entirely new viewpoints 
that have not been considered before by the community for addressing an important and crucial problem in EDA.
\section{Conclusion and Future Work} 
\label{sec:conclusion}
The field of EDA has to solve complex problems for IC design, often relying on heuristics. Quantum computers offer potential solutions through their optimization capabilities, yet research on leveraging them for EDA problems is limited. This paper explores the feasibility and potential of quantum computing for a typical EDA optimization problem, successfully executed on an IBM quantum computer and D-Wave's quantum annealing machines. Despite current qubit limitations, the presumably ongoing exponential growth in qubit availability suggests that quantum computing holds promise for EDA challenges. With problem size scaling linearly, quantum optimization techniques could provide effective solutions in the future.

Moving forward, our future endeavors encompass conducting additional analyses of this specific EDA problem, involving multiple executions on a quantum computer. Additionally, we aim to explore and analyze various other EDA problems like scheduling, placement and routing.
\section*{Acknowledgement}
We acknowledge the use of IBM Quantum services through the Fraunhofer Quantum Programme. The views expressed are those of the authors and do not reflect the official policy or position of IBM or the IBM Quantum team.

\bibliographystyle{ACM-Reference-Format}
\bibliography{ce}
\end{document}